\documentclass[useAMS,usenatbib,usegraphicx,referee]{mn2e}

\def\beq{\begin{equation}}
\def\eeq{\end{equation}}
\def\bey{\begin{eqnarray}}
\def\eey{\end{eqnarray}}

\def\yr{\,{\rm yr}}
\def\mas{\,{\rm mas}}
\def\masyr{\,{\rm mas}\,\yr^{-1}}

\def\kms{{\rm \,km\,s^{-1}}}
\def\kpc{\,{\rm {kpc}}}
\def\pc{\,{\rm {pc}}}

\def\vr{V_{\rm r}}
\def\dvr{\delta V_{\rm r}}
\def\mul{\mu_{\rm l}}
\def\vt{\overline{V_l}}

\def\araa{ARAA}
\def\apj{ApJ}
\def\aap{A\&A}
\def\aj{AJ}
\def\nat{Nature}
\def\mnras{MNRAS}

\def\thetabar{\theta_{\rm bar}}

\def\Dmin{D_{\rm min}}
\def\Dmax{D_{\rm max}}
\def\Ds{D_{\rm s}}

\def\eps@scaling{1.0}%
\newcommand\plottwo[2]{%
 \centering 
 \leavevmode 
 \columnwidth=.45\columnwidth
 \includegraphics[width={\eps@scaling\columnwidth}]{#1}%
 \hfil 
 \includegraphics[width={\eps@scaling\columnwidth}]{#2}%
}%

\begin{document}
\title{CONSTRAINING THE GALACTIC BAR PARAMETERS WITH RED CLUMP GIANTS}
\author[Mao \& Paczy\'nski]{
Shude Mao$^{1,2}$ \thanks{E-mail: smao@jb.man.ac.uk (SM);
bp@astro.princeton.edu (BP)} and Bohdan Paczy\'nski$^{2}$ \\
$^{1}$ Univ. of Manchester, Jodrell Bank Observatory,
  Macclesfield, Cheshire SK11 9DL, UK \\
$^{2}$ Princeton University Observatory, Princeton, NJ
08544-1001, USA
}

\date{Accepted . Received ; in original form }
\pagerange{\pageref{firstpage}--\pageref{lastpage}} \pubyear{2002}
\maketitle

\label{firstpage}

\begin{abstract}
Red clump giants in the Galactic bulge occupy a distinct
region in the colour magnitude diagram. They
have a very small spread in intrinsic luminosities and their
number counts have a well defined peak. 
We show that these characteristics can be used
to constrain the differences in the streaming motions of stars on
the near side and those on the far side. We propose two
methods to select two samples with one preferentially
on the near side and the other on the far side. In the first method,
we divide red clump giants into a bright sample
and a faint one; stars in the bright sample will be on average
more on the near side and vice versa. The second method
relies on the fact that lensed bulge stars lie preferentially on the far side
due to the enhanced lensing probability by the stars on the near
side and in the disk. If the radial streaming motion is $\approx 50\kms$,
we find the difference in the average radial velocity between
the bright and faint samples can reach $\approx 33\kms$ while
the corresponding difference is about $\approx 10\kms$ between
the lensed stars and all observed stars.
The difference in the average proper motion between the bright and faint samples
is about $\approx 1.6 \mas\yr^{-1}$ if there is a tangential
streaming motion of 100$\kms$; the corresponding shift between
the lensed stars and all observed stars is 
approximately $\approx 1\mas\yr^{-1}$. To observe the shifts in the radial
velocity and proper motion, roughly one hundred 
microlensing events, and/or bright/faint red clump giants,
need to be observed either spectroscopically or
astrometrically. The spectroscopic observations
can be performed efficiently using multi-object spectrographs 
already available. The proper motion signature
of microlensed objects can be studied using ground-based
telescopes and the Hubble Space Telescope.
These observations will provide 
strong constraints on the Galactic bar parameters.
\end{abstract}


\section{INTRODUCTION}

Many observational groups carried/are carrying out microlensing 
observations (e.g., MACHO: Alcock et al. 1993; OGLE: Udalski et al. 1993;
DUO: Alard \& Guibert 1997; MOA: Bond et al. 2001; EROS: Aubourg et al. 1993).
Over one thousand microlensing events in the local
group have been identified (e.g., Alcock et al. 2000; Wo\'zniak et al. 2001;
Derue et al. 2001; 
Alard \& Guibert 1997; Bond et al. 2001). In the coming years,
$\sim$ 1000 microlensing events are expected to be discovered
by the OGLE III\footnote{
http://www.astrouw.edu.pl/\~\,ogle/ogle3/ews/ews.html
} and other collaborations every year, many of them in
real-time. 

Data collected from microlensing experiments have very diverse applications
(for reviews see Paczy\'nski 1996; Gould 1996).
One of the most important applications is studying the Galactic structure.
There is strong evidence that the centre of Galaxy hosts a bar 
(e.g., de Vaucouleurs 1964; Blitz \& Spergel 1991;
Stanek et al. 1994, 1997; Kiraga \& Paczy\'nski 1994;
H\"afner et al. 2000 and references therein). However, the
parameters of the bar are not well-determined, including its mass,
size, and the motion of stars within it. 

Data from microlensing experiments 
provide several ways of probing the structure of the inner Galaxy. 
For example, the optical depth is roughly proportional to the total mass
of the bar while the event time scale distribution
probes the mass function and kinematics of the bar. Another
important method that was first explored by Stanek et al. (1997) uses
red clump giants (RCGs). These bulge stars occupy a distinct region
in the colour-magnitude diagram 
(see, e.g., Stanek et al. 2000 and references therein). They
have very small intrinsic widths in their luminosity function,
about 0.2\,mag for RCGs in the Galactic bulge
(see \S2 in Stanek et al. 1997; Paczy\'nski \& Stanek 1998).
The observed luminosity
function has a well-defined peak and an apparent width (see Fig.~5 in
Stanek et al. 1997). The apparent width depends
on both the intrinsic width and the spread caused by the 
radial depth of the bar. Stanek et al. (1997) used 
this dependence to constrain the Galactic bar axial ratios and orientation.

In this paper, we use the RCGs to
study the kinematics of stars in the Galactic bar. Stars on the
front side and far side may have different (tangential and
radial) streaming motions. We explore two effects
to constrain such motions. The first effect is based on the 
realisation that red clump stars at the bright slope
of the peak of the luminosity function
must be (on average) closer to us, i.e. on the near (front) side,
while the stars on the faint slope must be (on average)
on the far (back) side. Therefore, one can select bright and faint samples
of RCGs and examine their differences in the radial 
velocity and proper motion. The second effect we explore
is based on the well-known fact that
lensed bulge stars lie preferentially on the far side of the Galactic
bar due to the enhanced lensing probability by the stars on the near
side. This leads to several observable effects. For example, the
lensed RCGs
should be fainter compared with all observed red clump stars in the field
(Stanek 1995; see also Zhao 1999a, 1999b, 2000 for other 
effects such as extinction and reddening for microlensing events toward the
Large Magellanic Cloud). We will examine
systematic differences in the proper motions and radial 
velocities of the lensed RCGs
relative to all observed RCGs in the Galactic bar; a similar
bias in the radial velocity for lensing events toward
the Large Magellanic Cloud has been discussed by Zhao (1999b).

As this paper is the first
feasibility study, we do not 
attempt to construct a detailed (or even self-consistent)
model of the Galactic bulge. Our aim is to provide
an order-of-magnitude estimate of the kinematic biases in the hope of
motivating observations to be carried out, which may in
turn provide strong constraints on the 
the model of the inner Galaxy and incentives for their
refinement. The structure of the paper is as follows. In \S2,
we outline our model, and in \S 3 we present our main 
results.  We discuss observational issues
to detect these effects in \S 4.
For definiteness, we shall restrict ourselves to studying
the red clump stars in Baade's window ($l=1^\circ$, $b=-3.9^\circ$).

\section{MODEL}

To explore both effects that we discussed above for RCGs, 
we need to adopt a bulge model, which is 
rather uncertain; there are a number of published
analytical and numerical models of various
complexities (see Han \& Gould 1995 for a large selection of
analytical models; see Zhao et al. 1996,
H\"afner et al. 2000, Fux 2001 and references therein for numerical models).
For illustrative purposes, we will take
two simple analytical models in the literature for the Galactic bulge.

The first toy model we adopt 
is identical to the one used by Kiraga \& Paczy\'nski (1994).
The bulge mass density distribution is approximated
as axis-symmetric and is based on the study of Kent (1992). Note that
there is no bar in this simple model. The bulge velocity
distribution is assumed to be isotropic. Each of the three
velocity components follows a Gaussian distribution with a
dispersion of $100\kms$. In addition,
the bulge stars on the near side rotate with a velocity
of $+100\kms$ (in the same direction as the solar rotation)
while those on the far side rotate in the opposite sense (see
Kiraga \& Paczy\'nski for details). In this model,
the streaming motion is purely tangential. The second model we consider
is somewhat more realistic. It is based on the E2 bar model of
Dwek et al. (1995). For the following discussions, it is convenient to
establish a coordinate system centred on the bar with
$x, y$ and $z$ corresponding to  the major, intermediate and minor
axes. The minor axis coincides with the vertical direction while the major
axis has a viewing angle of $\thetabar=24^\circ$ from
the line of sight (for an illustration of the bar
geometry, see Figs. 6 and 11 in Stanek et al. 1997).
The axial ratios are taken to be $1:0.42:0.28$. The luminosity-weighted
velocity dispersions can be derived using the virial theorem
(Belokurov \& Evans 2001) as $\sigma_x=100\kms$,
$\sigma_y=80\kms$ and $\sigma_z=68\kms$, respectively.
In addition, the stars on the near side have a streaming motion
of $+50\kms$ parallel to the major axis,
while the stars on the far sides move in the opposite sense. Note that
the (rotational) sense
of the streaming motion is similar to that adopted by Kiraga \&
Paczy\'nski (1994), but a factor of two smaller in value.

The intrinsic luminosity function of red clump stars is assumed to follow
a Gaussian distribution with a width $\sigma_0 \approx 0.2\,{\rm
mag}$ (Stanek et al. 1997); there is presumably no difference in the luminosity
function for RCGs on the near and far sides. We assume that RCGs follow
the same bulge mass distribution over a distance range 
from $\Dmin<\Ds<\Dmax$; we take $\Dmin=0.5R_0$ and $\Dmax=1.5R_0$,
where $R_0=8\kpc$ is the distance to the Galactic centre.
Other choices of $\Dmin$ and $\Dmax$ only change our results slightly. 
Note that the effect of extinction and reddening can be minimised by
using the extinction-insensitive Wesenheit index, $W_{V-I}$
(written as $V_{V-I}$ in Paczy\'nski et al. 1994). 

RCGs are usually selected in the colour-magnitude diagram by requiring
their colours and magnitudes to be in a certain range (e.g., Fig. 1 in
Paczy\'nski \& Stanek 1998). Many red giants
occupy the same region in the colour-magnitude diagram. As the
luminosities of red giants span a broad range
they can be at either on the far side or on the near side
at any given (apparent) magnitude. 
Therefore, they dilute the kinematic differences 
between a bright RCG sample and a faint one
as their apparent brightness can not be used as an approximate
measure of distance. We model the number counts
of red giants as a linear function 
of the apparent magnitude (see eq. 1 in Paczy\'nski \& Stanek 1998).
According to the model, within one magnitude of the RCG peak, red 
giants contribute about 60\% of the total number. Therefore, red giants
substantially dilute the offset signals if they cannot
be separated from the red clump stars; we return to this
issue in the discussion.

For the offset in kinematics between lensed
and all observed RCGs, we need to calculate
the microlensing optical depth and event rate.
We consider both bulge self-lensing and lensing by the disc.
Therefore, we need to specify the disc mass distribution
and kinematics. As the offsets in the radial velocity and proper motion 
depend only on the {\it ratio} of the lensing frequency for stars on the near 
side and those on the far side, they
do not depend on the mass function as the mass cancels out in the ratio.

For the disc mass distribution, we adopt the familiar double exponential
form (Bahcall 1986)
\beq
\rho(r, z) = \rho_\odot \exp\left(-{r-R_0 \over 3.5\kpc} - {|z|
	\over 0.325\kpc}\right),
\eeq
where $(r, \phi, z)$ are the usual cylindrical coordinates (in units of
kpc),
and $\rho_\odot$ is the local disc density, for which we take 
a value of $0.06M_\odot\pc^{-3}$.  The velocity dispersions are taken to
be $\sigma_r=30\kms, \sigma_\phi=20\kms$, and $\sigma_z=20\kms$, which are
typical values in all disc models (e.g., Belokurov \& Evans 2002).
The disc lenses have a systematic rotation of $v_\phi=+220\kms$ at
all distances. In addition, we also account for the solar motion
relative to the local standard of rest (see e.g., Binney \& Merrifield 1998).

Obviously, the optical depth and event rate need to be
weighted properly by the source distance (see eqs. 5 and 11
in Kiraga \& Paczy\'nski 1994). For the lensing of RCGs,
we assume that they are bright enough to be seen with equal
probability at all distances
(corresponding to $\beta=0$ in the formalism of Kiraga \& Paczy\'nski
1994).

\section{RESULTS}

Monte Carlo simulations are performed to obtain our results; this
is the easiest way to incorporate different source populations and
selection strategies. In the following, we first 
discuss the offset in the radial velocity, $\vr$, and in the longitudinal
proper motion, $\mul$, between the faint and bright samples of RCGs,
and then discuss the offsets between the lensed and all observed RCGs.

\subsection{Offsets in $\vr$ and $\mul$ Between Faint and Bright Sub-samples of
RCGs}

As we discussed previously, the observed luminosity
function of RCGs has a well defined peak in their number
counts (see Fig.~5 in Stanek et al. 1997).
The RCGs on the brighter side of the peak
are (on average) on the front side of the bar and vice versa. 
We experimented with different ways of selecting `bright' and
`faint' samples in order to maximise their differences 
in the radial velocity and proper motion. We found that
the optimum `bright' sample is formed by stars that are brighter 
than the peak magnitude by 0.2 to 0.4 magnitude.
Similarly, the optimum `faint' sample is formed by stars fainter than
the peak magnitude by 0.2 to 0.4 magnitude. 
Notice that each sample also includes a background of red giant stars.
 
The shift in the radial velocity between the faint and bright
samples depends on the model. In the Kiraga \& Paczy\'nski model,
there is no radial streaming of stars, hence there is no shift in the
radial velocity as the stars on the near and far sides
have the same random kinematics.
This is different for the Dwek et al. model, where we have 
streaming motions of $\pm 50\kms$ mostly along the radial direction
as the bar is (by assumption) only $24^\circ$ from the line of sight.

\begin{figure*}
\plottwo{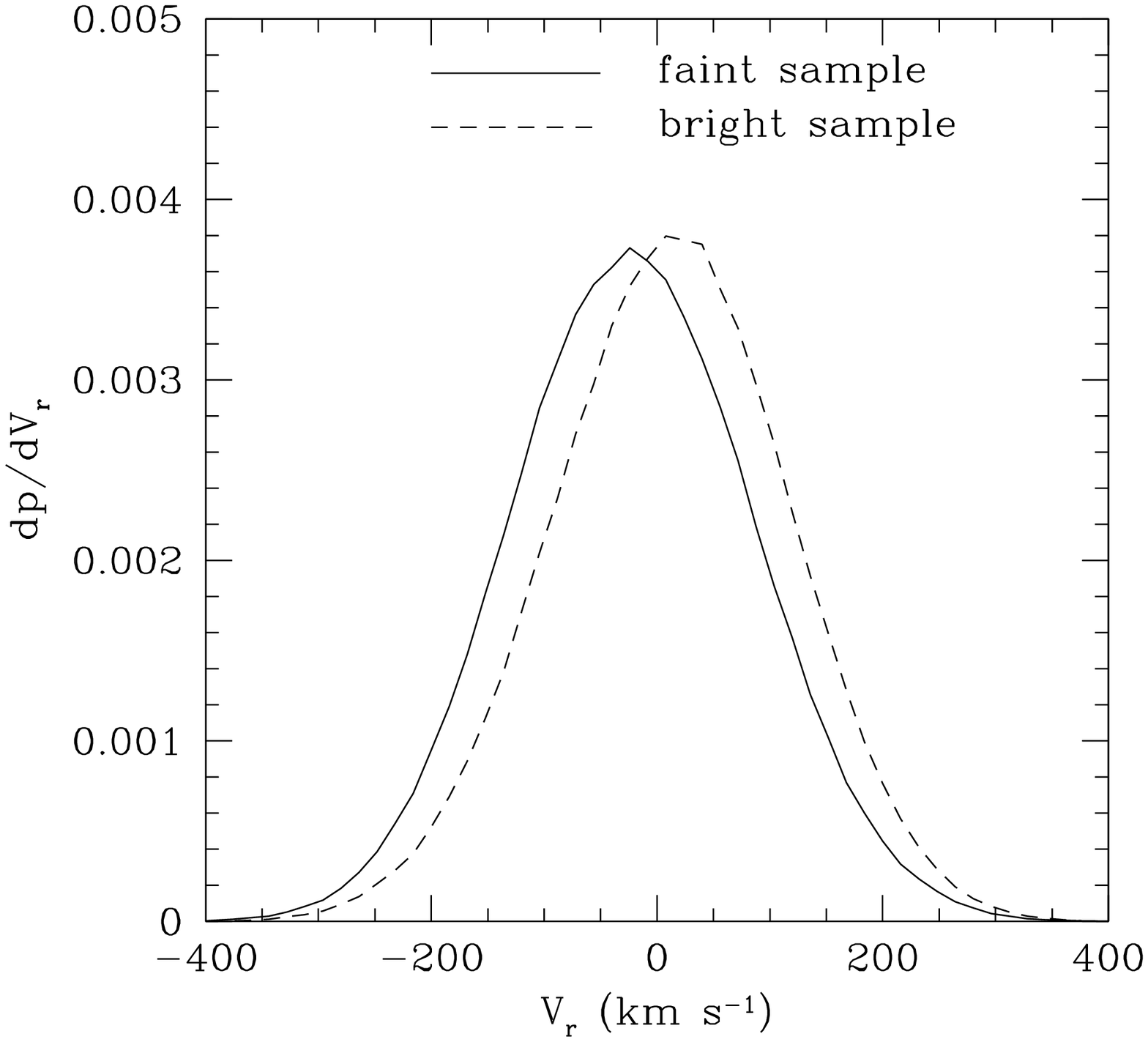}{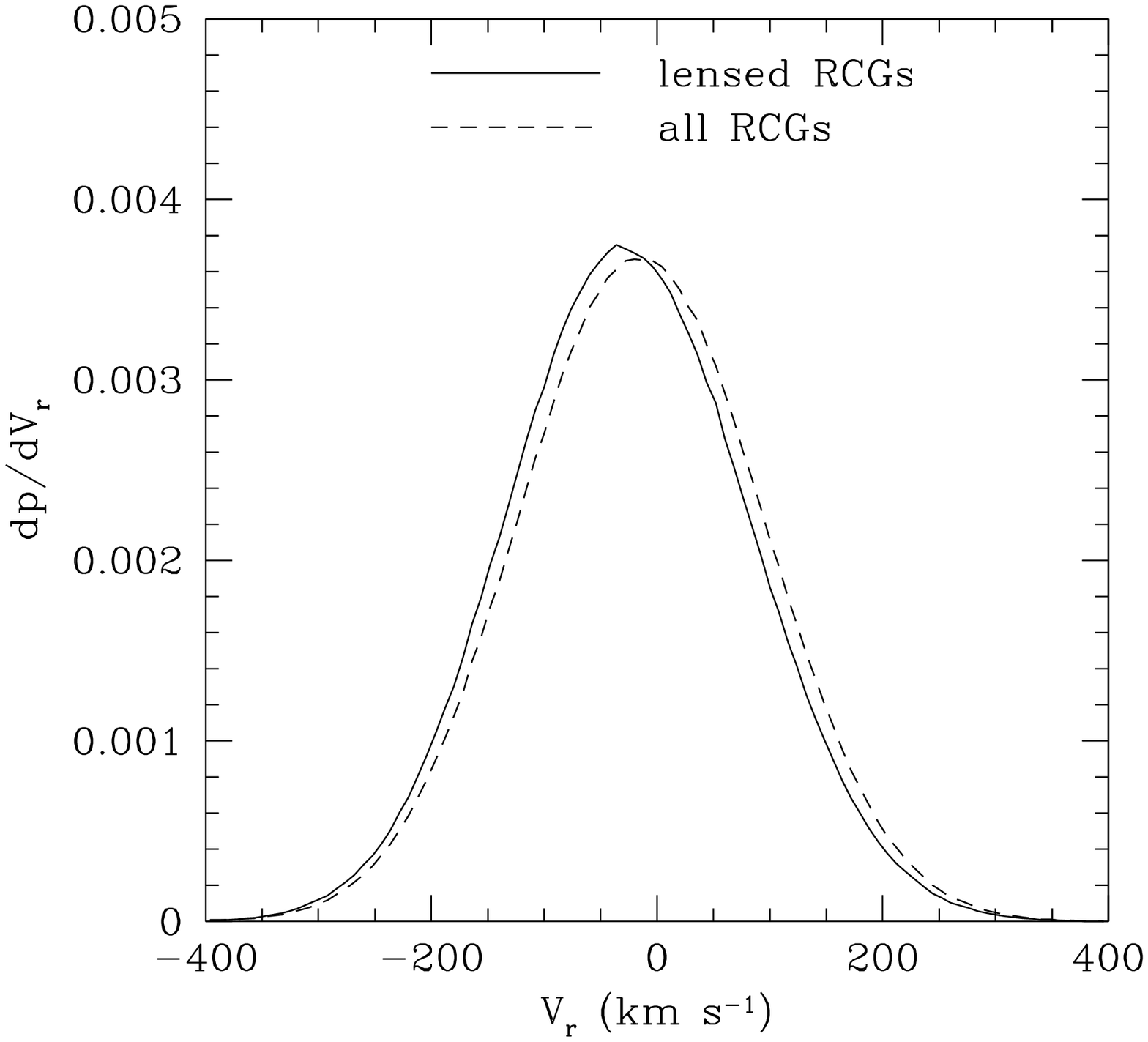}
\caption{Normalised differential probability
distributions of radial velocities. 
For the {\it left} panel, the solid (dashed) line shows 
the distribution for stars fainter (brighter) by 0.2 to 0.4 magnitude 
than the magnitude where the number counts of red
clump giants (RCGs) peak. 
The average shift between these two populations is about $33\kms$.
The {\it right} panel shows the distributions
for lensed (solid) and all observed (dashed) stars, respectively.
The average shift between these two populations is about $10\kms$.
In both panels, the Galactic bulge is
modelled as a bar with the major axis $24^\circ$ away from the line of
sight. The bulge sources have streaming motions parallel to
the major axis: stars on the near side move with
$+50\kms$ in the same sense of Galactic rotation while stars on the far
side move in the opposite sense.
\label{fig:vr}
}
\end{figure*}

The left panel in Fig.~\ref{fig:vr}
shows the distributions of radial velocity for the bright 
(dashed) and  faint (solid) samples, respectively. 
Both distributions are roughly Gaussian with a dispersion
of $\approx 110\kms$. The faint sample is shifted toward
more negative radial velocities, with an average shift
of $\approx 33\kms$. This shift can be understood qualitatively. The stars
on the near side have an average radial velocity
$(-10+50\cos\thetabar)\kms$,
where $-10\kms$ arises due to the solar motion ($+10\kms$) 
relative to the local standard of
rest. Analogously, stars in the faint sample are mostly on the
far side, and they have
an average radial velocity $(-10-50\cos\thetabar)\kms$.
As the faint sample includes more stars on the far side, hence
their radial velocities are more negative.
The average shift between the bright and faint samples 
is about one third of the velocity dispersion along the line of sight
($\sim 100\kms$). We return to the feasibility of observing
this shift in the discussion.

\begin{figure*}
\plottwo{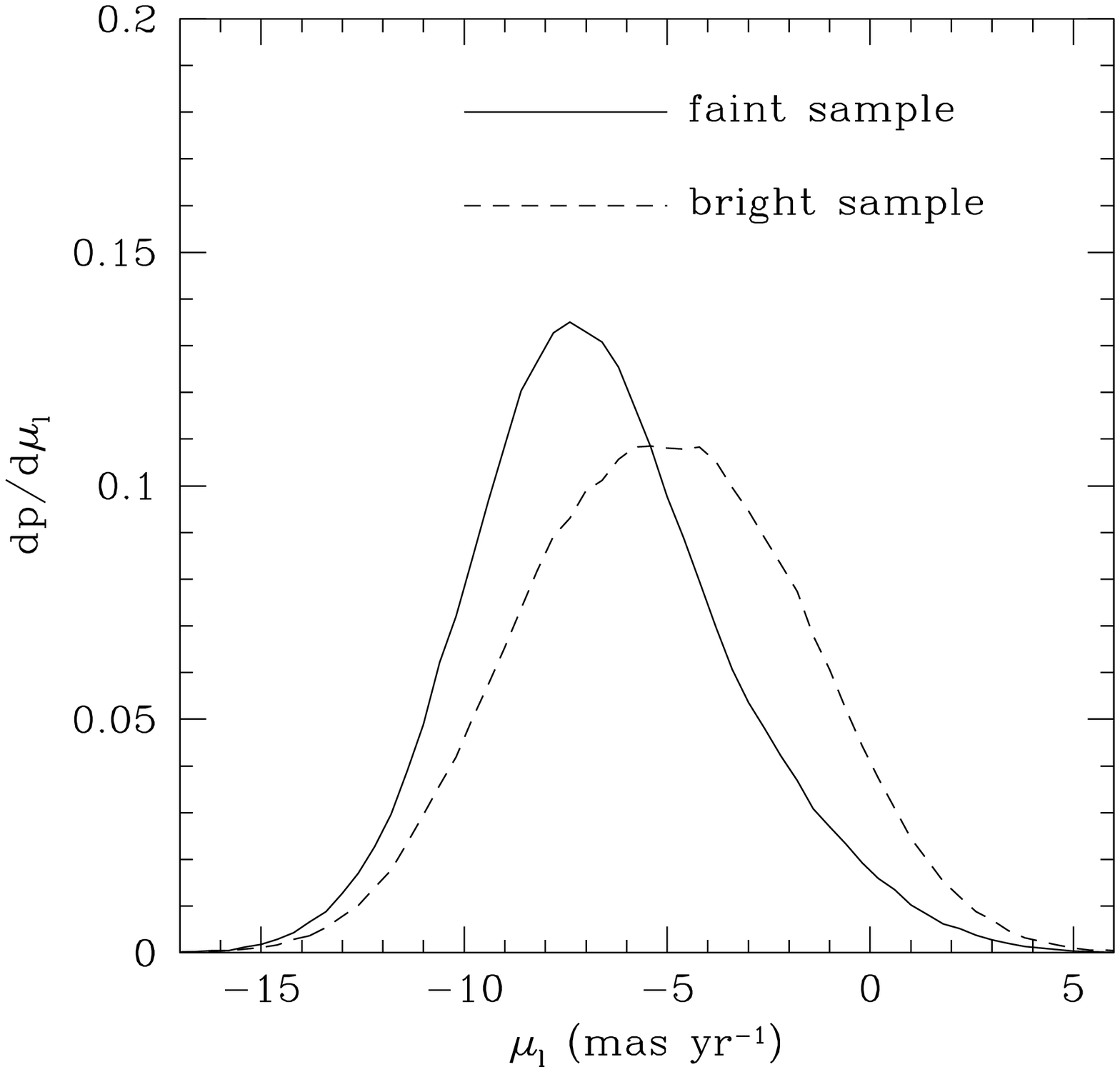}{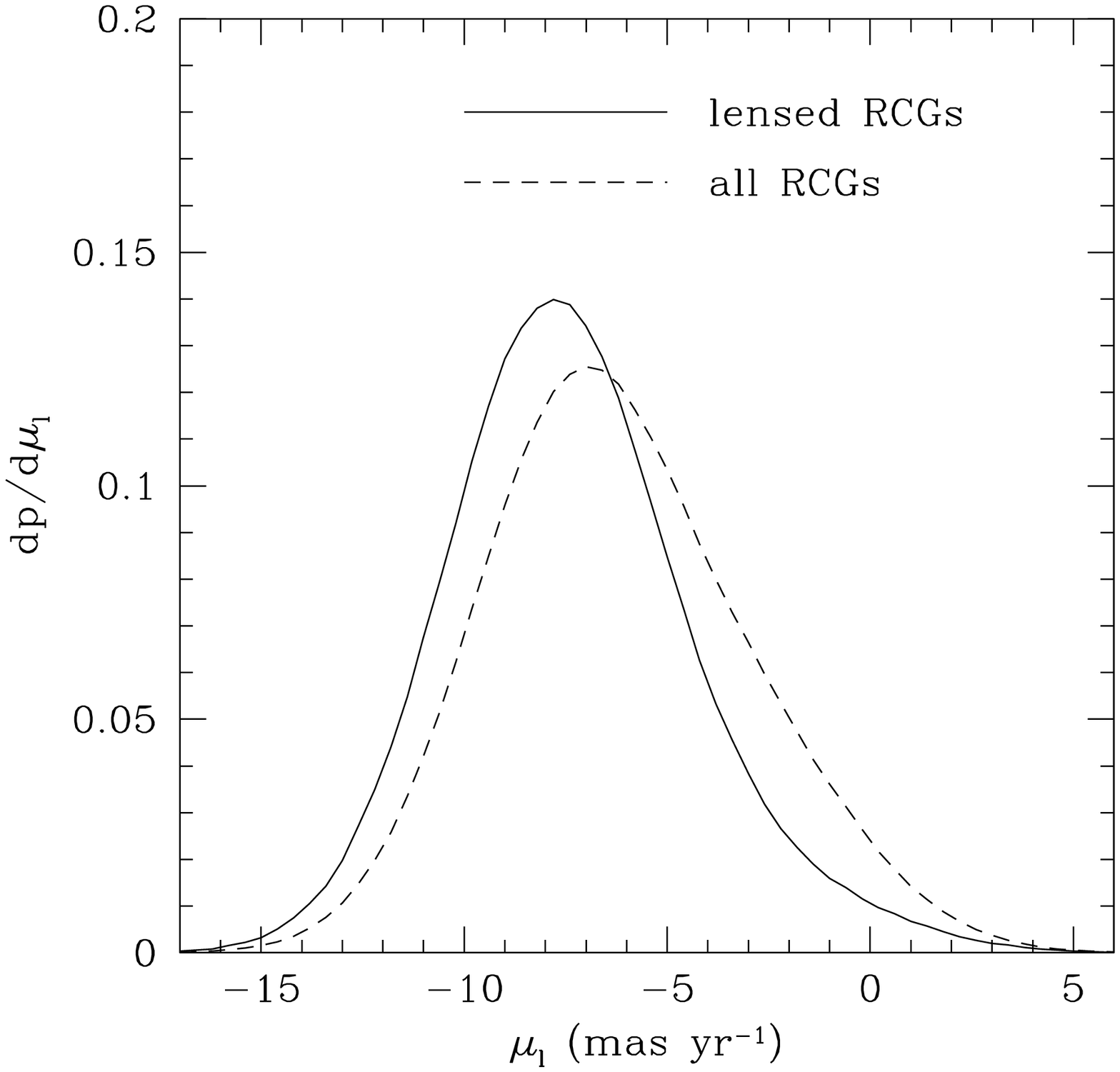}
\caption{Normalised differential probability
distributions of proper motions in the longitudinal direction
 in the Kiraga \& Paczy\'nski (1994) model. 
For the {\it left} panel, the solid (dashed) line shows 
the distribution for stars fainter (brighter) by 0.2 to 0.4 magnitude 
than the magnitude where the number counts of red
clump giants peak. On average, the fainter population
is shifted to a more negative proper motion by about $1.6\mas\yr^{-1}$.
For the {\it right} panel, the distributions for lensed and all observed RCGs
are shown as the solid and dashed lines, respectively. 
 The lensed population is shifted to a more negative proper
motion of about $1.2\mas\yr^{-1}$ on average. For both panels,
the bulge sources are assumed
to have a constant streaming motion perpendicular to the line of sight
of $100\kms$ but with opposite signs on the near and far sides of the 
Galactic centre.
\label{fig:mu}
}
\end{figure*}

As the streaming motion we considered is in the plane of the Galaxy,
we do not expect any shift in the latitudinal proper motion,
so we will only consider the shift in the longitudinal direction, $\mul$.
The shift in $\mul$ between the bright and faint
samples again depends on the
model. In the second model, the streaming motion is mostly radial, so
the proper motion shift is relatively small.
However, for the Kiraga \& Paczy\'nski model,
there is a tangential streaming motion of $+100\kms$ for stars
on the near side and  $-100\kms$ on the far side, and so the offset becomes
more apparent. The left panel in Fig.~\ref{fig:mu} shows
the distributions of proper motion for the bright (dashed)
and faint (solid) samples, respectively. The proper motions
for the faint sample is shifted toward a more negative value,
by an average amount of $1.6\masyr$; the dispersion for the faint
sample is also about 10\% smaller. To understand these trends,
let us recall that for a star with
velocity $v_l$ (in units of $\kms$) and at distance $\Ds$ (in units of
kpc) has a proper motion of $\mul  \approx 0.2 v_l/\Ds \masyr$.
In the {\it heliocentric} frame, stars on the far side
move with a streaming motion of $v_l=-300\kms$ while
stars on the near side move with $v_l=-100\kms$. As the faint sample
includes more stars from the far side, their proper motions
are therefore more negative relative to the bright sample.
The average shift is roughly proportional to the value of
the tangential streaming motion; it is about one half
of the dispersion in the proper motion
of approximately $3\mas\yr^{-1}$ for all observed stars.
The dispersion of the faint sample is smaller
because $\mul \propto \Ds^{-1}$ and
faint RCGs are on average at slightly larger distances.

\subsection{Offsets in $\vr$ and $\mul$ Between Lensed and All Observed RCGs}

The shift in the radial velocity between the lensed 
and all observed RCGs depends on the models. In the first
model where there is no radial streaming of stars,
there is no shift in the
radial velocity as the stars on the near and far sides
have the same random kinematics.
This is different for the second model, where we do have 
radial streaming motions.

The right panel in Fig.~\ref{fig:vr}
shows the distributions of radial velocity for the lensed
(solid) and all observed (dashed) RCGs, respectively. 
The curve predicted for all observed stars
has a negative mean value around $-10\kms$ partly 
due to the solar motion with respect to the local
standard of rest.  However, in this model, the stars
on the far side move with $-50\cos\thetabar\kms$ along the line of sight
and those on the near side move in the opposite direction. As the
optical depth and lensing event rate are higher for stars
on the far side, radial velocities of the lensed sources
are shifted toward more negative values.
The difference in the average radial velocity is about $\sim 10\kms$. 
This is smaller than the offset between the
bright and faint samples that we discussed in \S3.1. 
We found that this shift is roughly proportional to the value of
the radial streaming motion. In any case, the average shift
is fairly small compared with the velocity dispersion along the line of sight
($\sim 100\kms$).

The shift in the proper motion between the lensed
and all observed RCGs again depends on the
model. In the second model, the streaming motion is mostly radial, so
the shift in proper motion is relatively small.
However, for the first model,
where there is a tangential streaming motion of $+100\kms$ for stars
on the near side and  $-100\kms$ on the far side, the offset becomes
noticeable. The right panel in Fig.~\ref{fig:mu} shows
the distributions of proper motion for the lensed stars (solid)
and all observed stars (dashed), respectively. The curve for the
overall population has
an average proper motion of $\approx -5.8\mas\yr^{-1}$ due to the solar 
motion around the Galactic centre. The curve for the lensed population
is shifted toward more negative proper motions. This is
because stars on the far side move opposite to the solar rotation, and
as they have larger optical depths and are more frequently lensed, 
their proper motions are shifted
toward more negative longitudinal values. The average shift
in the proper motion is about $1.2\mas\yr^{-1}$, which is smaller
than its dispersion, about $3\masyr$. As expected,
this offset is roughly proportional to the value of
the tangential streaming motion.

\section{DISCUSSION}

In this paper we have studied how to use red clump giants (RCGs) 
to constrain the kinematics of stars in the Galactic bar. The basic
idea is that stars on the near and far sides may have different 
streaming motions, and if we can select two
samples with one preferentially
on the near side and the other on the far side, then 
they should show differences in the radial and tangential
streaming motions. We have examined two ways that we can select such samples.
The first method selects a bright sample and a faint one with
the peak magnitude of the number counts of RCGs as the approximate dividing line.
The RCGs in the faint sample is (on average) more on the far side of the
bar and vice versa. The second method relies on the fact that
stars on the far side are preferentially microlensed due
to the enhanced lensing probability 
by the stars on the near side and in the disc.  Hence the
lensed stars and all observed stars should behave differently in
kinematics.

We illustrate the kinematical differences 
in the context of two simple models of the
Galactic bulge. In the first (axis-symmetric) model,  the streaming motion
is purely tangential, while in the second (more realistic)
bar model the streaming motion is more or less
radial. In reality, the streaming motions in the Galactic bar may
have both a radial component and a tangential one, 
so our numbers should be taken as rough order-of-magnitude estimates.

We found at the difference in the radial velocity between the
bright and faint samples can reach  $\dvr \approx 33\kms$
for a radial streaming motion of $50\kms$. The difference
is substantially reduced by the population of red giants that
occupy the same part of the colour-magnitude diagram as the
RCGs, as they have a large span in luminosity, so a brighter
magnitude does not signal that a star is closer to us on average.
If there is an effective way of differentiating red giants
and red clump stars, then the differences in the radial velocity
and proper motion can be much larger. This can be, for example,
done by examining colours as 
RCGs occupy a slightly bluer part of the colour-magnitude
diagram. Spectroscopic observations may also ultimately provide
features that distinguish red giants and RCGs. For example, if
the contribution of red giants can be reduced by one half, then
the shift in the proper motions can be
as high as $2.3\masyr$, while that for the 
radial velocity can reach $50\kms$.
In comparison, we find that 
the shift between the lensed stars and overall population of stars is
more modest, about $10\kms$ for a radial streaming motion of $50\kms$. 

The velocity dispersion along the line of sight
is about $\sigma\approx 100\kms$, 
so in order to see this shift at the
$2\sigma$ level, from Poisson statistics,
we only need to obtain the radial velocity for
about $(2 \times \sigma/\dvr)^2 \approx 100$ stars even
for $\dvr=20\kms$. The required number is moderate, but
is within reach of multi-object spectrographs
available on many large telescopes. For example, the FORS1 instrument on VLT
has 19 slits that can be efficiently used to derive the radial
velocities of lensed stars and field stars simultaneously. Other
instruments such as 2dF \footnote{http://www.aao.gov.au/2df/} which 
can take up to 400 spectra simultaneously should also be explored. We
note that for RCGs, blending is not a severe problem as they
are so bright, so their radial velocities can be measured without much
difficulty. In any case, we only require an accuracy of
$\sim 10-20\kms$ per star, since the dispersion in the radial velocity for the
bulge stars is of the order of $100\kms$.

For the shift in the proper motions, we find that
a value of $1.6\mas\yr^{-1}$ between the bright and faint
samples of RCGs for a tangential streaming
motion $\vt =100\kms$. The shift scales roughly linearly with
$\vt$. The difference is
about $1\mas\yr^{-1}$ between the lensed and overall population
of bright stars. These shifts are not much smaller than
the dispersion in the proper motion, approximately
$3\mas\yr^{-1}$. The shifts can be detected with HST
or even ground-based instruments. 
To do this, only a modest number of RCGs
need to be monitored to measure the relative shifts in proper motions.
The relative proper motion may be feasible to detect
even from the ground. Soszy\'nski et al. (2002)
demonstrated that proper motions
as small as $4\mas\yr^{-1}$ can be measured
with OGLE II data. This accuracy is achieved
through a combination of
the new difference image analysis software (Eyer \& Wo\'zniak 2001)
and a large number of data points for any given star, which reduces the
astrometric error through Poisson statistics.
The camera system of OGLE III yields much better
images than that of OGLE II as the latter operates in the drift scan
mode and produced somewhat worse point spread functions.
Therefore it seems possible (although still to be demonstrated)
that proper motions
of many red clump stars, including lensed ones, can
be obtained using data from
microlensing surveys (e.g., OGLE III), requiring
no extra observing resources.

In this paper we have
only explored the shifts in the radial velocity and proper motion
in Baade's window. Clearly there must be a spatial dependence of these shifts.
The OGLE III collaboration is currently monitoring many fields around
the Galactic centre, covering from $-7^\circ$ to $6^\circ$ in latitude
and $-12^\circ$ to $12^\circ$ in longitude
\footnote{For detailed field coordinates, see
http://www.astrouw.edu.pl/\~\,ogle/ogle3/ews/gb\_ews.jpg}. 
Red clump stars and microlensing events
will become available over large areas, and so
it will be very interesting to detect the spatial variation of the shifts
in the radial velocity and proper motion. Such observations will likely
provide strong constraints on the Galactic bar parameters, and
stimulate further theoretical efforts to
build a better and self-consistent model of the inner Galaxy. Finally,
we point out that if future astrometric missions,
SIM and GAIA\footnote{http://sim.jpl.nasa.gov;
http://astro.estec.esa.nl/GAIA/}, perform as planned,
they will directly measure the radial depth of the 
galactic bar by determining parallax's for a number of stars.  This
will make our suggestion obsolete, but not sooner than in a decade.

We thank Martin Smith, David Spergel, and Hongsheng Zhao for helpful
discussions. S.M. acknowledges travel support by Princeton University.
This project was supported by the NSF grant AST-1206213, and the NASA grant
NAG5-12212 and funds for proposal \#09518 provided by NASA through a grant from
the Space Telescope Science Institute, which is operated by the Association of
Universities for Research in Astronomy, Inc., under NASA contract NAS5-26555. 

%
%

\newpage

\newpage

\bsp

\label{lastpage}

\end{document}